\documentclass[fleqn,usenatbib]{mnras}

\usepackage{amsmath}
\usepackage[varg]{txfonts}

\usepackage[T1]{fontenc}

\DeclareRobustCommand{\VAN}[3]{#2}
\let\VANthebibliography\thebibliography
\def\thebibliography{\DeclareRobustCommand{\VAN}[3]{##3}\VANthebibliography}

\usepackage{graphicx}	
\usepackage{amsmath}	

\usepackage{natbib}
\usepackage{enumitem}
\usepackage[flushleft]{threeparttable}
\usepackage{colortbl}
\usepackage[normalem]{ulem}
\usepackage{color}
\usepackage{bm} 
\usepackage{gensymb}
\definecolor{darkblue}{rgb}{0.0,0.0,0.8}
\definecolor{darkred}{rgb}{0.75,0.,0.25}
\definecolor{darkorange}{rgb}{1,0.3,0.0}
\definecolor{darkgreen}{rgb}{0.0,0.6,0.0}
\definecolor{darkpurple}{rgb}{0.8,0.,0.9}
\definecolor{brown}{rgb}{0.65,.16,0.16}
\definecolor{grey}{rgb}{0.4,0.5,0.6}
\definecolor{white}{rgb}{1,1,1}
\definecolor{trolleygrey}{rgb}{0.5, 0.5, 0.5}
\definecolor{lavender}{rgb}{0.835,0.812,0.969}
\definecolor{pastelorange}{rgb}{0.99,0.92,0.82}
\definecolor{pastelblue}{rgb}{0.85,0.93,0.99}

\newcommand{\gaia}{{\sl Gaia}}

\newcommand{\probP}{\text{I\kern-0.15em P}}
\newcommand{\diff}{{\rm d}}

\title[Central stellar depletion in Draco]{Central stellar depletion in the Draco dwarf from Subaru photometry}

\author[Vitral et al.]{Eduardo Vitral$^{1}$\thanks{Email: eduardo.vitral@roe.ac.uk}\thanks{Royal Society Newton International Fellow},
Kyosuke S. Sato$^{2,3}$,
Jorge Pe\~narrubia$^{1,4,5}$,
Sakurako Okamoto$^{2,3}$
and Veronika Dornan$^{1}$
\\
$^1$Institute for Astronomy, University of Edinburgh, Royal Observatory, Blackford Hill, Edinburgh EH9 3HJ, UK\\
$^2$Astronomical Science Program, The Graduate University for Advanced Studies, SOKENDAI, 2-21-1 Osawa, Mitaka, Tokyo 181-8588, Japan\\
$^3$National Astronomical Observatory of Japan, 2-21-1 Osawa, Mitaka, Tokyo 181-8588, Japan\\
$^4$Institute of Corpuscular Physics (IFIC), CSIC–Universitat de València, 46980 Paterna, Valencia, Spain\\
$^5$VALER, Calle Mayor, 83, 1, 12001 Castellón de la Plana, Spain
}

\date{Accepted XXX. Received YYY; in original form ZZZ}

\pubyear{\the\year{}}

\begin{document}
\label{firstpage}
\pagerange{\pageref{firstpage}--\pageref{lastpage}}
\maketitle

\begin{abstract}
Dark matter subhaloes are a fundamental prediction of the standard cosmological model, but their low-mass population remains difficult to constrain observationally. Recent work has shown that interactions between subhaloes and dwarf spheroidal galaxies can induce large-scale, out-of-equilibrium fluctuations in their stellar surface density fields, identifiable through Fourier-based statistics. We apply this formalism to deep Subaru photometry of the Draco dwarf, extending previous tests to $\gtrsim\mathcal{O}(10^{4})$ resolved stars. We use artificial-star tests to assign completeness weights to individual stars and incorporate them into a discrete likelihood-based framework. Relative to a cored axisymmetric Plummer model, Draco shows radial corrugations that appear in the filtered normalised residuals as a central stellar depletion surrounded by a ring-like excess. The same feature appears in Fourier space as a low-frequency peak with a signal-to-noise ratio ${\rm S/N}=6.0$. Independently, fitting the surface density $\Sigma$ with an $\alpha\beta\gamma$ model yields ${\rm d}\Sigma/{\rm d}R>0$ at $>99.9\%$ confidence -- i.e., a profile that decreases towards the centre. Although these signatures cannot yet be regarded as robust evidence for dark subhaloes, they suggest that central stellar depletion in dwarf galaxies may reflect genuine physical structure rather than crowding alone.
\end{abstract}



\section{Introduction} \label{sec: intro}

In the standard $\Lambda$ cold dark matter ($\Lambda$CDM) picture, galaxies form within extended dark matter haloes that host a hierarchy of smaller, gravitationally bound subhaloes \citep[see][for a review on this topic, with references therein]{Zavala&Frenk19}. The abundance and internal properties of this substructure encode fundamental information on the nature of dark matter \citep{Colin+02,Bullock&Boylan-Kolchin17,Yang+23,Chiang+25}, yet most subhaloes are expected to remain entirely dark and therefore difficult to detect directly \citep[e.g.][]{Benitez-Llambay&Frenk20}. This has motivated a wide range of indirect probes \citep{Mao&Schneider98,Zechlin+12,Erkal+16,Bovy+17, Drlica-Wagner+19,Nadler+21,Ballard+24,Delos25,Enzi+25,Cao+25,Tajalli+25}, but current constraints still leave much of the low-mass subhalo population weakly tested \citep[e.g.][]{Bonaca+19,Banik+21}. Dwarf spheroidal galaxies offer an especially attractive laboratory in this context: they are strongly dark matter dominated \citep{Pryor&Kormendy&Kormendy90,Pace+24}, have relatively simple baryonic histories \citep{Savino+25}, and provide a setting where the dynamical imprint of dark perturbers may be isolated more cleanly than in systems with more complex stellar structure \citep{Penarrubia+24,Penarrubia+25}.

Motivated by this, \citeauthor*{Vitral+26-subhalos} (\citeyear{Vitral+26-subhalos}, hereafter V26) developed a Fourier-based formalism to search for subhalo-induced perturbations in the stellar surface density field of dwarf spheroidals, using $N$-body realisations from \cite{Penarrubia+25}. In that framework, repeated interactions with dark subhaloes drive a gradual, self-similar expansion of the stellar component within the host dwarf galaxy potential. In addition, the presence of dark subhaloes leads to subtle departures from smooth equilibrium, such as large-scale density corrugations whose signatures can be more straightforwardly identified in Fourier space. The method is particularly sensitive to the most massive surviving subhaloes in the host halo, and its practical detectability depends not only on the underlying subhalo population but also on the number of available stellar tracers. This makes nearby dwarfs with large resolved stellar samples especially valuable targets for an observational implementation of the method.

Among the Milky Way satellites, Draco is a particularly compelling test case: it is one of the most dark matter-dominated classical dwarf spheroidals \citep[e.g.][]{Read+18,Hayashi+20}, its star formation shut down long ago \citep{Aparicio+01}, and it appears to have been unaffected by tides \citep[][]{Odenkirchen+01,Segall+07}. Recent dynamical analyses further indicate that Draco is consistent with a cusped dark matter profile, in good agreement with the $\Lambda$CDM expectation \citep[e.g.][]{Vitral+24,Yang+25,Pascale+26}. In that sense, Draco is close to the kind of simple, cusp-dominated system explored in the idealized models of \cite{Penarrubia+25}, making it a natural galaxy in which to test whether the Fourier formalism can recover physically meaningful structure from real data.

A key practical difficulty is that these signals are subtle and require large stellar samples to be distinguished robustly from shot noise (V26). In this respect, deep wide-field Subaru imaging offers an important opportunity to move beyond the tracer counts accessible from \gaia's shallow depth regime, or from DES fields strongly affected by crowding \citep[e.g.][]{Wang+19-fornax6}. Datasets taken by Hyper Suprime-Cam (HSC) on the Subaru telescope extend the resolved stellar map of Draco to fainter magnitudes and larger radii while the next generation of facilities, such as Roman, are still being tested \citep{2026arXiv260221280H}.
In this Letter, we therefore use the HSC point-source catalogue from \cite{2026arXiv260916476S}
to apply the Fourier-based formalism to Draco’s stellar distribution and assess whether its resolved structure shows evidence for the kind of fluctuations expected from interactions with dark subhaloes.

The remainder of this Letter is organized as follows. In Section~\ref{sec: methods}, we summarize the main aspects of the Subaru/HSC data and describe the methods used to implement the Fourier-based formalism of V26, highlighting the relevant assumptions and caveats. In Section~\ref{sec: results}, we present and discuss the main results of the analysis. Finally, Section~\ref{sec: conclusion} summarizes our conclusions.

\section{Data \& Methods} \label{sec: methods}

\begin{figure}
\centering
\includegraphics[width=0.8\hsize]{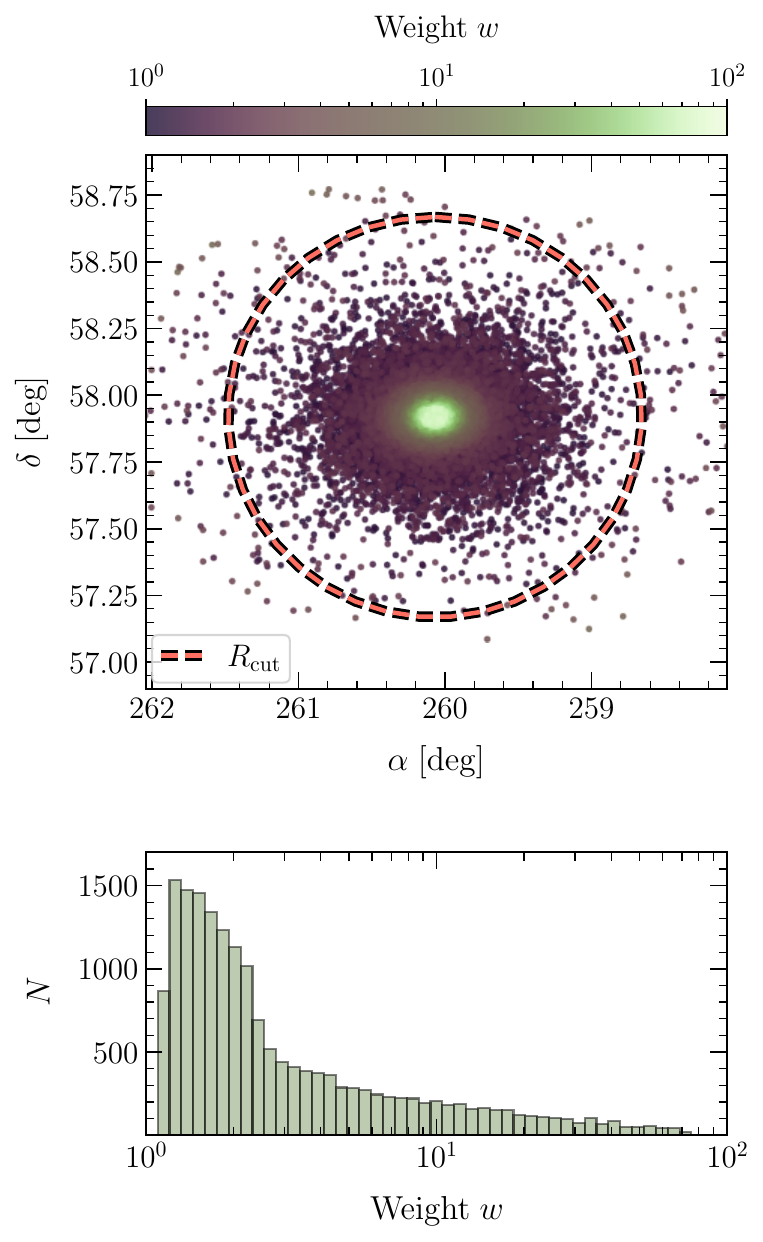}
\caption{\textit{Completeness weights:}
The \textbf{top} panel shows the sky-projected scatter distribution of the Subaru/HSC stellar data. The red dashed circle marks the region within which stars were included in our fits and Fourier-based analysis. Stars are colour-coded according to the completeness weights derived from the artificial-star tests described in Section~\ref{sec: methods}. Darker purple tones indicate stars with higher completeness (smaller weights), whereas lighter green tones correspond to stars requiring larger completeness corrections, predominantly in denser, more crowded regions. The \textbf{bottom} panel shows the distribution of these completeness weights using logarithmically spaced bins.}
\label{fig: weight-map}
\end{figure}

\subsection{Subaru/HSC data}
\label{ssec: subaru-reduction}

We utilise Subaru/HSC imaging of the Draco dSph, covering the system out to a projected radius of approximately $1$~kpc. The photometric data were reduced using \textsc{hscPipe~8.5.3}, with the full reduction procedure described by \citet{2026arXiv260916476S}. The resulting catalogue reaches approximately three magnitudes below Draco's old main-sequence turn-off.

We estimated the detection completeness independently in the $g$- and $i_2$-bands using artificial-star tests. Rather than adopting a binned two-dimensional completeness map, which could introduce artificial spatial discontinuities, we measured the completeness as a function of magnitude within $3\arcmin$-wide elliptical annuli, adopting $\epsilon=0.266$ and ${\rm P.A.}=92\fdg47$ \citep{2026arXiv260916476S}. Within each annulus, the magnitude dependence of the completeness was fitted with a sigmoid function \citep{2016ApJ...833..167M}. The corresponding sigmoid parameters were then linearly interpolated as a function of elliptical radius, yielding continuous completeness functions $C_g(g,R_{\rm ell})$ and $C_{i_2}(i_2,R_{\rm ell})$.

Each star was assigned a completeness-correction weight,
\begin{equation}
w
=
\left[
C_g(g,R_{\rm ell}) \, C_{i_2}(i_2,R_{\rm ell})
\right]^{-1},
\end{equation}
such that $w=1$ corresponds to full completeness, while larger weights account for decreasing completeness at a given position and magnitude. 
Figure~\ref{fig: weight-map} shows the stars used in our analysis within a $45\farcm0$ radius, colour-coded by their assigned completeness weight, $w$. The corresponding distribution of weights is shown in the bottom panel for reference. As expected, the weights increase towards the more crowded central regions of the system, where stellar incompleteness is more significant.

In this Letter, we restrict our analysis to old main-sequence (MS) stars, thereby avoiding the substantially higher contamination affecting the red giant branch; the motivation and methodology underlying this selection are described in detail by \citet{2026arXiv260916476S}. The resulting catalogue contains 17,524 stars, of which 17,301 lie within our adopted $45\farcm0$ aperture and have valid completeness weights. Accounting for these weights gives a completeness-corrected total count of $\sum_i w_i = 93,503$ stars, rounded to the nearest integer.

This observed sample of $\mathcal{O}(10^4)$ stars, corresponding to a completeness-corrected count of $\mathcal{O}(10^5)$, is beyond an order of magnitude larger than the $\mathcal{O}(10^3)$ \gaia\ counterpart considered in Table~1 of V26, while the completeness correction accounts for the estimated number of stars missed by the observations.
As discussed by V26, a larger stellar sample reduces shot noise in the reconstructed density field and can therefore improve the contrast of density fluctuations relative to an adopted smooth baseline. The depth of the Subaru/HSC catalogue thus provides a substantially denser sampling of Draco's stellar distribution for the density-fluctuation analysis developed below.

\subsection{Application of Fourier-based formalism}
\label{ssec: fourier-formalism}

\begin{figure*}
\centering
\includegraphics[width=\hsize]{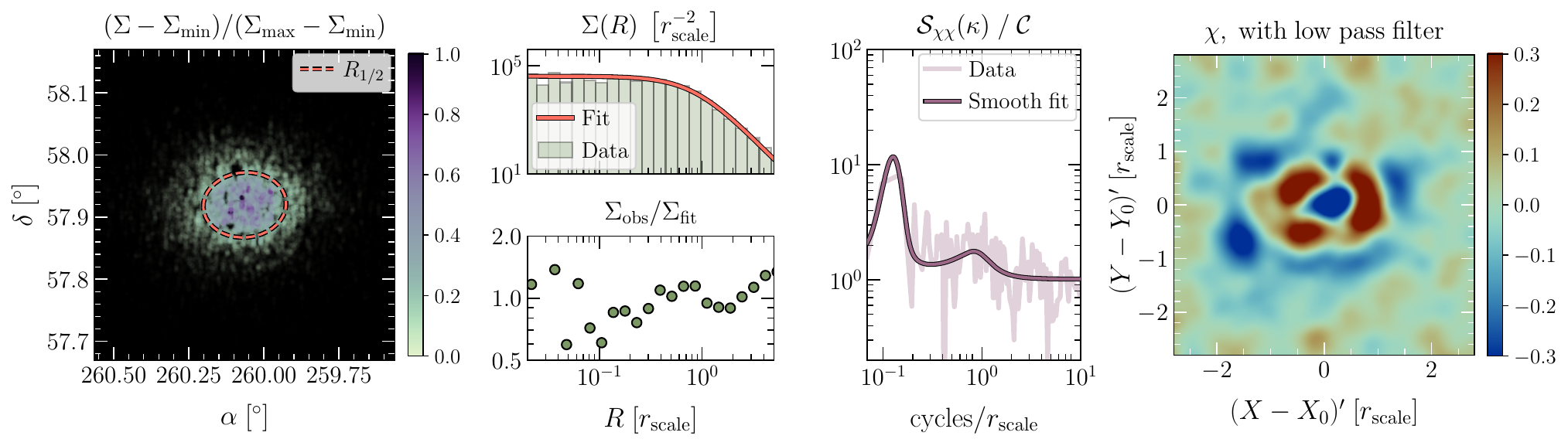}
\caption{\textit{Application to Draco:}
We apply the methodology of \protect\cite{Vitral+26-subhalos} to Subaru/HSC data of the Draco dwarf galaxy, which contain $\mathcal{O}(10^{4})$ member stars. The \textbf{first column} shows the observed projected Cartesian density map, where both colour and transparency scale with the local relative stellar surface density; the red dashed ellipse marks the fitted projected half-number ellipse. The \textbf{second column} is divided into two rows. The \textit{top} row shows the radial surface density profile (green) as a function of projected distance from the galaxy centre, expressed in units of the scale radius defined in Eq.~A1 of \protect\cite{Vitral+26-subhalos}, together with the best-fitting axisymmetric Plummer model (red). The \textit{bottom} row shows the corresponding radial residuals, defined as the ratio between the observed and fitted $\Sigma(R)$ profiles, and reveals a faint corrugated pattern. Both binned profiles in this column are shown solely as goodness-of-fit diagnostics and do not enter the fitting procedure, which is performed directly on the individual stellar positions. The \textbf{third} and \textbf{fourth columns} present the azimuthally averaged one-dimensional power spectrum of the $\chi$ field and the corresponding low-pass filtered $\chi$ map, respectively. The residual power spectrum shows an excess of low-frequency power, reflected in the filtered $\chi$ map as large-scale, ring-like fluctuations similar to those recovered in the controlled subhalo experiments of \protect\cite{Penarrubia+25} and \protect\cite{Vitral+26-subhalos}, where blue indicates depletion with respect to the smooth Plummer baseline, and red indicates stellar excess. This makes Draco a particularly compelling case for deeper follow-up.}
\label{fig: full-chi-plot}
\end{figure*}

We follow the procedure described in Sections~3, 4.1, and 4.2 of V26 to construct Fourier-filtered $\chi$ maps and account for random Poisson fluctuations. Here, we summarize its main ingredients and highlight the only modification relevant to the present application: unlike in V26, each star $j$ is assigned an individual weight, $w_j \geq 1$, derived from our data-reduction procedure and artificial-star tests.

At its core, the method quantifies local departures of the observed stellar counts from a smooth reference baseline. Following \cite{Penarrubia+25} and V26, we adopt an axisymmetric Plummer \citep{Plummer1911} profile as the reference smooth stellar distribution, characterized by a central constant-density core. This profile has also been shown to provide satisfactory fits to Draco in \cite{Vitral+24}. 
As in V26, the centroid of the stellar distribution is treated as a free parameter of the Plummer fit, together with its scale radius, ellipticity, and position angle. The centre adopted throughout the subsequent analysis therefore corresponds to the best-fitting photometric centroid rather than being fixed to a literature value.

For each spatial cell in a two-dimensional grid,\footnote{The technical details on how this grid is constructed follow the exact same description provided in V26.} departures from this smooth model are quantified through the statistic
\begin{equation} \label{eq: chi-measure-poisson}
\chi = \frac{N_{\rm obs} - N_{\rm P}}{\sqrt{N_{\rm P}}} \ ,
\end{equation}
where $N_{\rm obs}$ and $N_{\rm P}$ denote the observed and model-predicted counts in each spatial cell, respectively. In the present completeness-corrected application, $N_{\rm obs}$ is replaced by the corresponding weighted count, $N_{\rm obs}=\sum_j w_j$, and the model normalization is adjusted consistently using the total weighted number of stars. We otherwise retain the Poisson normalization of the original V26 statistic, $\sqrt{N_{\rm P}}$.

The corresponding likelihood is therefore modified to incorporate these weights:
\begin{equation}
    \mathcal{L} \equiv \prod_{\rm j} \frac{{\diff \probP} }{\diff \mathcal{S}} = \prod_{\rm j} \left( \Sigma_{\rm ilop} / N_{\rm tot} + \sum_{\rm i}^{N} \Sigma_{\mathrm{sys}, \, i} / N_{\rm tot} \right)^{w_{j}} \ .
    \label{eq:likelihood}
\end{equation}
These weights are propagated consistently throughout the analysis. In particular, they enter the surface density goodness-of-fit diagnostics, the normalization of the total number of stars, and the computation of the $\chi$ statistic itself through weighted histograms. In the limit $w_j = 1$ for all stars, one recovers the expressions from the original formalism.

We then analyze the resulting $\chi$ field in Fourier space through its azimuthally averaged power spectrum, which we model as
\begin{equation} \label{eq: power-spectrum}
\mathcal{S}_{\chi \chi, \, {\rm fit}}(\kappa, \{\theta\}) = \mathcal{C} + \sum_{i=1}^{n} f_i \, V(\kappa, \, \mu_i, \ q_i \, \delta_i, \ (1 - q_i) \, \delta_i) \ ,
\end{equation}
where $\kappa$ is the radial spatial frequency, $\mathcal{C}$ is the high-frequency noise floor, and $V$ denotes a Voigt profile. For each component $i$, the parameter set $\{ \theta \}_{i}$ is composed of: $f_i$, which sets the flux of the feature, a central frequency $\mu_i$, a characteristic scale $\delta_i$, and the mixing parameter $q_i$ controlling the relative Gaussian and Lorentzian contributions. In this notation, the Gaussian width is $\sigma_i = q_i \delta_i$, while the Lorentzian scale length is $\gamma_i = (1-q_i)\delta_i$. As discussed in V26, the dominant signal is typically associated with the primary Voigt component, whose fitted position and width trace the characteristic spatial scales of the fluctuations.

We quantify the significance of this feature through its signal-to-noise ratio, defining the signal as the primary Voigt flux normalized by the high-frequency noise floor, $(f_1/\mathcal{C})$. We estimate this normalized flux across bootstrap realizations of the power-spectrum fit and derive its uncertainty, $\sigma_{(f_1/\mathcal{C})}$, from the resulting bootstrap distribution. The detection significance is then defined as ${\rm S/N}=(f_1/\mathcal{C})/\sigma_{(f_1/\mathcal{C})}$. This statistic quantifies how significantly the primary low-frequency feature is recovered relative to its fitting uncertainty.

In practice, we use this weighted framework to derive $\chi$ maps for the Draco data, determine the Voigt parametrization that best reproduces the azimuthally averaged observed power spectrum, and apply a Butterworth \citep{Butterworth30} filter\footnote{As in the previous work, we adopt $n = 4$ and $\kappa_{\rm Cutoff} = 1~{\rm cycle}/ r_{\rm scale}$ in the relation $B(\kappa) = 1/\sqrt{1 + (\kappa / \kappa_{\rm Cutoff})^{2n}}$ (cf. eq.~8 of V26).} to suppress high-frequency spatial noise. This yields a figure directly analogous to Fig.~11 of V26, where the same methodology was applied to \gaia\ data, albeit with $w_j = 1$. We discuss the resulting maps and their implications in the following sections.

\section{Results \& Discussion} \label{sec: results}

We show the outcome of our Fourier-based analysis in Figure~\ref{fig: full-chi-plot}. As in V26, the leftmost panel presents the observed projected Cartesian density map, where both colour and opacity scale with the local relative stellar surface density and the red dashed curve marks the fitted projected half-number ellipse. The second column is divided into two rows: the \textit{top} row shows the radial surface density profile (green) as a function of projected distance from the galaxy centre, expressed in units of the scale radius defined in Eq.~A1 of V26, together with the best-fitting axisymmetric Plummer model (red), while the \textit{bottom} row shows the corresponding radial residuals, defined as the ratio between the observed and fitted $\Sigma(R)$ profiles. The third and fourth columns present the azimuthally averaged one-dimensional power spectrum of the $\chi$ field and the corresponding low-pass-filtered $\chi$ map, respectively.
Interestingly, the radial ratio between the observed and fitted $\Sigma(R)$ profiles exhibits a corrugated pattern that differs somewhat from those predicted in V26 for subhalo-induced perturbations, most notably through an excess of stars in the outskirts ($R \gtrsim 2~r_{\rm scale}$) relative to the Plummer model.

In any case, the inner depletion that is common to both analyses is accompanied by a Voigt peak in the azimuthally averaged one-dimensional power spectrum of the $\chi$ field, detected with a signal-to-noise ratio of ${\rm S/N}=6.0$ (see definition in Section~\ref{ssec: fourier-formalism}).
After the corresponding Butterworth filtering, the final panel reveals, as anticipated from the previous diagnostics, a central depletion region in blue surrounded by a ring-like stellar excess in red. 
The presence of a centrally decreasing surface density is also supported by an independent check based on an $\alpha\beta\gamma$ \cite{Zhao96} profile fit,\footnote{This profile is fitted directly to the individual stellar positions using a weighted likelihood such as Eq.~\ref{eq:likelihood}, adopting the corresponding axisymmetric analytical form from \cite{Zhao96}.} which yields $\gamma = 0.36 \pm 0.03$, corresponding to $>99.9\%$ confidence for a declining central surface density profile. This result is also consistent with the recent analysis of \citet{Walker+26}, who found Draco's density profile to be statistically consistent, at the $1\sigma$ level, with a central depletion, despite masking the innermost regions of the galaxy in their analysis.

The interpretation of these results follows closely the discussion in V26. Even if our artificial-star tests provide a reliable description of the spatial completeness of the stellar sample, the measured stellar depletion in Draco cannot at present be robustly attributed to the cumulative impact of dark subhaloes alone, appealing though that possibility may be. The main limitation is that alternative mechanisms remain insufficiently tested. For example, it is still unclear whether baryonic feedback, or alternative dark matter models, could generate central stellar depletion on the same spatial scales and frequencies associated with subhalo-driven perturbations. Further work is therefore needed to disentangle distinct out-of-equilibrium signatures and establish more secure diagnostics based on complementary data. 
That said, progress is being made in this direction: in particular, \cite{Penarrubia&Nadler26} recently showed that stellar tidal stripping can halt, and even reverse, the gravothermal expansion described in \cite{Penarrubia+25}, thus suggesting that the signatures observed here are \textit{unlikely} to be driven by tides. This interpretation is further supported by the fact that stellar stripping is expected to begin only after most of the dark matter halo envelope has already been lost to tides \citep{Penarrubia+08,Vitral&Boldrini22}. Moreover, repeated searches for tidal debris around Draco have so far been unsuccessful \citep{Odenkirchen+01,Segall+07}, while its orbit inferred from \textit{Gaia} proper motions has a pericentre sufficiently large that substantial tidal mass loss is not expected \citep{Pace+22,Battaglia+22}.

In any case, our results suggest that observable central stellar depletion in dwarf spheroidal galaxies may be more common than previously assumed, calling for greater care when modelling surface-density profiles, particularly in analyses that mask central regions to avoid possible crowding effects. The results of V26 already indicate that such depletion may arise from genuine physical mechanisms rather than observational limitations alone, while approaches based on artificial-star tests, such as the one adopted here, can help quantify and mitigate residual crowding biases. 
This result may also have broader implications for dwarf-galaxy mass modelling, particularly for equilibrium analyses: several studies have argued that equilibrium orbital configurations become increasingly restricted as the stellar density profile becomes shallower within cuspy potentials \citep{An&Evans06,Ciotti&Morganti09,Ciotti&Morganti10b,Sanchez-Almeida+24-axi,Sanchez-Almeida+24-sph}. 
We stress that this does not necessarily imply that Draco itself is out of equilibrium -- indeed, its measured tangential anisotropy \citep{Vitral+24} allows a broader range of equilibrium configurations including shallower density profiles, while its non-spherical shape \citep{Walker+26} further weakens arguments based on more restrictive geometries. More generally, however, the observed central depletion could still be indicative of disequilibrium and motivates further scrutiny of the equilibrium assumptions commonly adopted when modelling dwarf spheroidal galaxies \citep[e.g.,][]{Read+21}.

\section{Conclusion} \label{sec: conclusion}

We have analysed state-of-the-art Subaru/HSC data for the Draco dwarf spheroidal galaxy. After using artificial-star tests to constrain the level of spatial incompleteness in the data, we assigned corresponding stellar weights to the sample and incorporated them into the fitting routine of V26. This consisted of fitting an axisymmetric Plummer profile to the projected stellar distribution and using Fourier-based statistics to distinguish coherent low-frequency structure from small-scale statistical fluctuations, thereby quantifying departures from a smooth baseline at low spatial frequencies.

This approach yields results qualitatively similar to the subhalo-induced signatures predicted in that work. In particular, we detect radial corrugations relative to a baseline Plummer profile. In the filtered $\chi$ map, where $\chi$ quantifies local departures from the smooth model, these appear as a central stellar depletion surrounded by a ring-like stellar excess. The same feature is recovered in Fourier space as a Voigt peak at $\kappa \approx 0.8~{\rm cycles}/r_{\rm scale}$ in the azimuthally averaged power spectrum of the residual $\chi$ field.

Given the early stage of development in this area, however, these signatures cannot yet be regarded as bona fide evidence for dark subhaloes in Draco. Establishing such a connection will require determining whether other mechanisms, including mergers and alternative dark matter models, can produce comparable spatial fluctuations. On the observational side, an important test will be to apply the method to datasets for which central crowding and the associated completeness corrections are less severe than in the Subaru sample considered here. Deep space-based imaging will be particularly valuable in this respect, as the absence of atmospheric seeing and the higher angular resolution should improve both stellar completeness and the separation of individual stars from compact background galaxies.

Nevertheless, if our artificial-star tests provide a reliable description of the spatial completeness, the central depletion recovered here suggests, as was also found for Fornax in V26, that such features need not arise solely from observational crowding, but may instead reflect genuine physical structure. This possibility calls for greater care when masking the inner regions of dwarf galaxies in surface-density profile fits \citep[see][for a discussion of this topic]{Walker+26}, while motivating further theoretical work to distinguish subhalo-induced fluctuations from those generated by other perturbing mechanisms.

Ultimately, more decisive tests will require combining wide-field, high-precision photometry from facilities such as the \textit{Nancy Grace Roman Space Telescope} and \textit{Euclid} with increasingly realistic simulations able to resolve dwarf galaxies with $\mathcal{O}(10^{5})$ stars. Complementary kinematic diagnostics offer an additional means of testing the subhalo scenario, including kinematic misalignments and low-amplitude rotation predicted in our subhalo experiments \citep{Vitral+26-rotation}, for which observational evidence has also been reported in Draco \citep{Vitral+24,Pascale+26}. Combining these photometric, theoretical, and kinematic constraints should provide a substantially stronger basis for assessing the role of dark subhaloes in shaping out-of-equilibrium stellar structure.

\section*{Acknowledgements}

EV acknowledges funding from the Royal Society, under the Newton International Fellowship programme (NIF\textbackslash R1\textbackslash 241973).
KS acknowledges funding from JST SPRING (JSPS), Japan Grant Number JPMJSP2104.
SO acknowledges support in part from the JSPS Grant-in-Aid for Scientific Research (JP25K01047) and JSPS Core-to-Core Program (grant number: JPJSCCA20260002).

The following software and tools were used extensively during the writing of this manuscript: {\sc Python} \citep{VanRossum09}, 
{\sc BALRoGO} \citep{Vitral21},
{\sc Scipy} \citep{Jones+01},
{\sc Numpy} \citep{vanderWalt11},
{\sc Matplotlib} \citep{Hunter07},
{\sc Visual Studio Code} \citep{microsoft2026vscode},
{\sc ChatGPT} \citep{openai2026chatgpt},
and the Scientific color maps from \cite*{Crameri+20}.
\section*{Data Availability}

The data that support the plots within this paper and other findings of this study are available from the corresponding author upon reasonable request.



\bibliographystyle{mnras}
\bibliography{src} 




\appendix



\bsp	
\label{lastpage}
\end{document}